\documentclass[a4paper,11pt]{article}
\usepackage{jheppub}
\usepackage{latexsym,amsmath,amsfonts,amssymb}
\usepackage{epsfig,graphics,graphicx,pgf,pstricks}
\usepackage[utf8]{inputenc}

\newcommand{\id}{{\bf 1}}

\newcommand{\be}{\begin{equation}}
\newcommand{\ee}{\end{equation}}

\newcommand{\Z}{{\mathbb Z}}

\newcommand{\new} {\operatorname{new}}
\newcommand{\old} {\operatorname{old}}

\newcommand{\im} {\operatorname{Im}}

\newcommand{\vev}[1] {\left<#1\right>}

\begin{document}

\title{A simple approach towards the sign problem using path optimisation}

\author{Francis Bursa$^{(1)}$, Michael Kroyter$^{(2)}$}
\affiliation{$^{(1)}$ School of Physics and Astronomy\\
University of Glasgow\\
Kelvin Building, University Avenue, Glasgow\\
G12 8QQ United Kingdom
\ \\
$^{(2)}$ Department of Sciences\\
Holon Institute of Technology (HIT)\\
52 Golomb St., Holon 5810201, Israel}
\emailAdd{francis@semichrome.net}
\emailAdd{michaelkro@hit.ac.il}

\abstract{
We suggest an approach for simulating theories with a sign problem that relies on optimisation of
complex integration contours that are not restricted to lie along Lefschetz thimbles.
To that end we consider the toy model of a one-dimensional Bose gas with chemical potential.
We identify the main contribution to the sign problem in this case as coming from a nearest neighbour interaction and 
approximately cancel it by an explicit deformation of the integration contour.
We extend the obtained expressions to more general ones, depending on a small set of parameters.
We find the optimal values of these parameters on a small lattice and study their range of validity.
We also identify precursors for the onset of the sign problem.
A fast method of evaluating the Jacobian related to the contour deformation is proposed and its numerical stability is
examined. For a particular choice of lattice parameters, we find that our approach increases the lattice size at which the sign
problem becomes serious from $L \approx 32$ to $L \approx 700$.
The efficient evaluation of the Jacobian ($O(L)$ for a sweep) 
results in running times that are of the order of a few minutes on a standard laptop.
}

\maketitle

\section{Introduction}
\label{sec:intro}

The sign problem poses a challenge for a lattice simulation of many physical theories,
ranging from widely-studied systems such as QCD and other theories with a finite chemical
potential~\cite{deForcrand:2010ys,Aarts:2015tyj}, through the simulation~\cite{Medina:2017xbn}
of PT-symmetric theories~\cite{Bender:1998ke,Bender:2007nj}, to the simulations~\cite{Bursa:2014oza}
of more exotic systems such as string field theory~\cite{Witten:1986cc}.
It also appears in the simulations of various condensed matter theories as well as in other systems.
In general, one faces a challenge whenever $e^{-S}$ is not a positive quantity.
In such a case $e^{-S}$ cannot be identified as a (non-normalized) probability density and the
Metropolis algorithm~\cite{Metropolis:1953am} cannot be applied without a modification. A simple modification is reweighting.
However, for fast oscillations of the phase of the action the computational cost of evaluation using reweighting grows
exponentially with the lattice volume.
The sign problem is this exponential behaviour.

Several methods have been introduced in order to circumvent the sign problem: One could attempt to expand the problematic part of the integrand
in a Taylor expansion such that all terms in the expansion involve only integrals with a constant phase~\cite{Gavai:2003mf}.
While this method works for some systems, it introduces an error in the estimate of observables on top of the standard statistical error.
Also, one could not expect it to be useful for large values of the parameters used in the expansion, e.g. the chemical potential.
Another approach is to analytically continue the problematic parameters to values that lead to no sign problem and then continue
back~\cite{deForcrand:2002hgr,DElia:2002tig}. While this method could also be used in some cases, numerical analytical continuation can be
quite challenging and the approach suffers from problems similar to those of the Taylor expansion approach.
A third approach is the complex Langevin method~\cite{Aarts:2013lcm}. Here, stochastic dynamics is used for the calculation of observables and there is
no reference to $e^{-S}$ as a probability density. Hence, the sign problem is avoided. Nonetheless, this method also has its
limitations~\cite{Aarts:2009uq,Aarts:2011ax,Nagata:2016vkn}.

An important ingredient of the complex Langevin method is the complexification of the degrees of freedom.
Complexification can be used in other ways as well:
One could attempt to avoid the phase oscillation causing the sign problem by using (the multi-dimensional form of)
Cauchy's integral theorem for deforming the original integration contour to one without (or at least with less) phase oscillations.
When using Cauchy's theorem, one must take care not to pass any singularities of the integrand and to deform the asymptotic
integration range only when it vanishes fast enough, e.g. using Jordan's lemma.
For most physical theories the analytical continuation of the action is regular. Hence, the first issue is of no concern for these theories.
The asymptotic behaviour of the integrand, on the other hand, can become singular for specific contours. For potentials
that behave asymptotically as $V\simeq\phi^n$ with $N$ variables ($N$ equals the product of field components by the number of lattice points)
there are generically $(n-1)^N$ different homology classes of integration contours~\cite{Guralnik:2007rx,Witten:2010cx}.
One should therefore be careful not to deform the contour away from the original homology class.

A possible prescription for the contour deformation is to use Lefschetz thimbles~\cite{Witten:2010cx}, which also give
the steepest descent of the (real part of the) measure. The implementation of Lefschetz thimbles in lattice simulations
was introduced in~\cite{Cristoforetti:2012su}.
Despite the success of this approach, it is also not without faults:
\begin{enumerate}
\item While it is known that Lefschetz thimbles are manifolds, explicit expressions defining these manifolds are absent.
This leads to complicated and expensive algorithms for verifying that the integration contour does not leave the thimble.
\item While in some cases only a single thimble contribution is relevant in the continuum limit~\cite{Cristoforetti:2012su},
in other cases one needs many thimbles. Since the mean phase factor,
i.e. $\vev{e^{i\im(S)}}_{PQ}$, the mean value of the phase in the phase quenched ensemble, can differ among
different thimbles, this could lead to reemergence of the sign problem as a ``global sign problem'', especially in light of the fact
that the number of homology classes (the number of independent thimbles) goes to infinity in the continuum limit.
\item Lefschetz thimbles are constructed in a way that keeps the imaginary part of the classical action constant.
However, when working with thimbles the integration measure changes. This leads to the
``residual sign problem''.
\item For any given lattice there is a different set of thimbles. One should therefore identify the thimbles
as well as their contribution to the desired cohomology class independently for every lattice size, lattice spacing, mass, etc.
This becomes more and more complicated as the lattice size is increased.
\end{enumerate}
The above issues can be summarized by noting that Lefschetz thimbles improve, but do not completely resolve
the sign problem,
while reducing the running cost from exponential to $O(V^4)$.
In some cases the computational cost of a thimble simulation can be further decreased~\cite{Alexandru:2016lsn}.
Nonetheless, it is still advisable to look for improvements and alternatives to the Lefschetz thimble method.
Indeed, generalisations of the thimble method have been proposed, in which the above mentioned issues are addressed
by choosing a single integration contour as an approximation to the sum of all the relevant
thimbles~\cite{Alexandru:2015xva,Alexandru:2016ejd,Bedaque:2017epw}.

In this work, we propose a method for a different deformation of the integration contour.
We do not attempt to approximate the thimbles. Thimbles are defined by a
gradient flow, which defines a real $N$-dimensional submanifold of the complex $N$-dimensional space.
On the other hand, the requirement of a vanishing imaginary part puts only one real constraint and hence
its solution is a $(2N-1)$-dimensional real manifold. We intend to choose an $N$-dimensional subspace of
this manifold, or an approximation to such a subspace, in order to obtain a proper integration cycle.
This might appear to be too arbitrary as compared to a cycle defined by a gradient flow.
However, there is much arbitrariness also in the definition of a gradient flow: In order to define
the gradient operator, a metric should be introduced in the complex $N$-dimensional space. It is possible
to choose this metric to be the flat metric, but this choice is not canonical~\cite{Witten:2010cx}.
The only relevant attribute of this metric is its compatibility with the complex structure, which
guarantees that the imaginary part is constant along the flow, i.e. the metric must be
K\"ahler~\cite{Witten:2010cx}.
While not all the manifolds with constant imaginary part and appropriate boundary conditions can be obtained
by varying the metric (also, some
metric variations lead to a reparametrisation rather than a change of the obtained manifold),
some can be obtained this way. This illustrates that even ``the thimble'' is not a uniquely defined object.
Moreover, an integration cycle which is not the thimble can be chosen
to be a single contour that already takes into account the change in the integration measure.
Hence, with a proper choice of integration
contour, both the residual sign problem and the global sign problem could be avoided.

For illustrating our method we consider the specific case of a Bose
gas at a finite chemical potential. This model is often used for the purpose of examining new methods for
dealing with the sign problem, e.g.~\cite{Aarts:2008wh,Cristoforetti:2012su}.
On the lattice (in lattice units) the action of the model is,
\begin{align}
\label{action}
S&=\sum_k \bigg(\Big(d+\frac{m^2}{2}\Big)(u_k^2+v_k^2)+\lambda\frac{(u_k^2+v_k^2)^2}{4}\\
\nonumber
  &-\sum_{\nu=1}^{d-1} \Big(u_k u_{k+\hat{\nu}}+v_k v_{k+\hat{\nu}}\Big)-\cosh(\mu)(u_k u_{k+\hat{0}}+v_k v_{k+\hat{0}})
	+i \sinh(\mu)(u_k v_{k+\hat{0}}-v_k u_{k+\hat{0}})
\bigg).
\end{align}
Here, $u_k,v_k$ are respectively the real and imaginary parts of a single complex field,
\begin{equation}
\Phi_k=u_k+iv_k\,,
\end{equation}
the index $k$ runs over all lattice sites and $d$ is the dimensionality of space-time.
While the signature is Euclidean, the ``time'' direction is special, since the interactions
related to it differ from those of the other coordinates due to the presence of the chemical potential $\mu$.
Moreover, the terms related to the chemical potential are the ones responsible for the
imaginary part of the action.
The action enjoys a $U(1)$ symmetry, even for a non-zero chemical potential.

Rescaling the fields by a factor of $\sqrt{\frac{2d+m^2}{\lambda}}$ the action becomes,
\begin{align}
\label{dDimAction}
&S=\frac{1}{\lambda\alpha^2}\sum_k \!\bigg(\!\frac{u_k^2+v_k^2}{2}+\frac{(u_k^2+v_k^2)^2}{4}\\
\nonumber
  &-\alpha \Big(\sum_{\nu=1}^{d-1}(u_k u_{k+\hat{\nu}}+v_k v_{k+\hat{\nu}})+\cosh(\mu)(u_k u_{k+\hat{0}}+v_k v_{k+\hat{0}})
	+i \sinh(\mu)(v_k u_{k+\hat{0}}-u_k v_{k+\hat{0}})
\Big)\!
\bigg).
\end{align}
where we defined
\begin{equation}
\label{alphaDef}
\alpha\equiv \frac{1}{2d+m^2}\,.
\end{equation}
For fixed $d$, this constant ranges from $\alpha=0$, which corresponds to the infinite mass limit to $\alpha=\frac{1}{2d}$,
which corresponds to the zero mass limit.
For small values of the chemical potential the vacuum of the theory is in an unbroken phase of the $U(1)$ symmetry,
while a spontaneously broken phase is obtained for
\begin{equation}
\cosh(\mu)>1+\frac{m^2}{2}\,.
\end{equation}
In the continuum limit the above equation takes the simpler form $\mu>m$.

\begin{figure}[ht]
\begin{center}
  \includegraphics[width=4in]{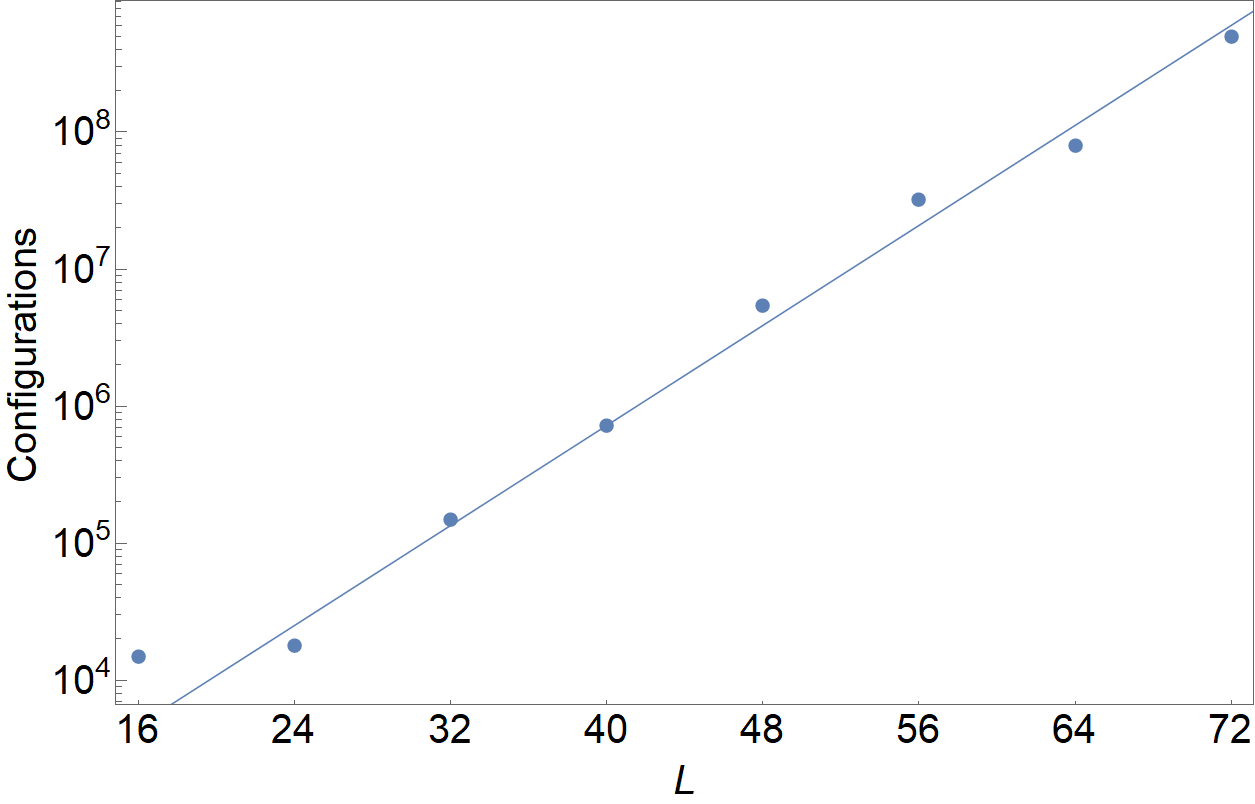}
\end{center}
\caption{The number of configurations needed for retaining a constant relative precision as a function of lattice size
for an unmodified contour grows exponentially as $e^{0.21L}$. The sign problem is evident.}
\label{fig:signProblem}
\end{figure}

For concreteness we concentrate in the rest of the paper on $d=1$.
In this case the first term in the second line of~(\ref{dDimAction}) is absent.
Assuming periodic boundary conditions and $L$ lattice points one can write
\begin{align}
\label{S1d}
S=\frac{1}{\lambda\alpha^2} \sum_{k=1}^L \bigg(&\frac{u_k^2+v_k^2}{2}+\frac{(u_k^2+v_k^2)^2}{4}\\
\nonumber
  &-\alpha \Big(\cosh(\mu)(u_k u_{k+1}+v_k v_{k+1})
	+i \sinh(\mu)(v_k u_{k+1}-u_k v_{k+1})
\Big)
\bigg).
\end{align}
As long as $\mu$ is small enough, taking the integration contour
to be the product of the real $u_k$ and $v_k$ axes would not lead to a significant phase factor.
As $\mu$ is increased the phase variations along this integration contour become  larger and larger
and the sign problem appears.
To illustrate this (well known) situation we attempted to simulate the system with the standard integration contour
on lattices of size $16,24,32,...,72$. For simplicity we set, as we do for all the simulations in this paper, $\lambda=m=\mu=1$
and estimated the number of configurations that should be sampled in order to
obtain a standard error of $\vev{S}$ that would be about 2\% of $\vev{S}$. Since this is only for
an illustrative purpose we were content with identifying a number of configurations that leads to a relative
error in the range of $1.6\%-2.4\%$. We find that while for $16$ and $24$ lattice points we need roughly the same
number of configurations, for larger lattices the required number of configurations grows fast.
We performed a least squares fit for the logarithm of the number of configurations as a function of lattice size
for the range $L=24...72$ and found that the number of configurations needed grows as $e^{0.21L}$.
We plot the number of configurations needed together with the fit found in
fig.~\ref{fig:signProblem}.
Assuming this exponential behaviour continues, we can estimate that the running time (on a standard laptop) with no contour
modification needed for the stated precision for $L=96$ would already be of the order of a couple of months.
This is significantly longer than the running time with contour deformations studied below,
for which the precision is also much better.

In light of the above we look for adequate contour deformations. We write the complexified fields as
\begin{equation}
u_k=x_k+i y_k\,,\qquad v_k=\xi_k+i\zeta_k\,,
\end{equation}
and look for a parametrisation of the contour of the form
\begin{equation}
\label{contourForm}
y_k=y_k(\{x_m,\xi_m\})\,,\qquad\zeta_k=\zeta_k(\{x_m,\xi_m\})\,,
\end{equation}
that is, we intend to use the real parts of the variables as the space over which we integrate.
This is not the most general form of a deformed contour: We do not allow contours that return to previous values of
$\{x_m,\xi_m\}$.
However, as we already stressed, there is very large freedom in choosing the integration contour
and our choice of parametrisation does not limit us. Finding an integration contour amounts to solving a single equation
\begin{equation}
\label{imS0}
\im(S)=0\,,
\end{equation}
in $2L$ variables. One could get to the conclusion that this is a trivial task. However, the solution must
be well defined and free from singularities in the whole space spanned by the $x_k$ and $\xi_k$, since otherwise
the ``contour'' would not be continuous and Cauchy's theorem would not hold.
Moreover, as stated above, care must be taken to ensure that the asymptotic behaviour of the contour is the desired one.
While it is still possible to allow the contour to approach other asymptotic regions before returning and heading to the correct one,
this should only be allowed as long as $e^{-S}$ is absolutely integrable along such paths.

How should one choose an integration contour?
We suggest to follow an old and important principle of physics:
Choose a contour that is as simple as possible but not simpler than that.
We should now decide upon our criteria for ``simplicity'' and upon the general considerations that should guide us
in our search after good integration contours:
\begin{enumerate}
\item Explicit and functionally simple form: We would like to have expressions that can easily be implemented and whose
computational cost is small.
\item Approximate expressions: The imaginary part of the action does not have to vanish exactly as in~(\ref{imS0}).
A small variation of the phase that does not lead to a severe sign problem can well be tolerated. When possible,
we can trade the accuracy of~(\ref{imS0}) for simplicity of the expressions.
\item Include the Jacobian if needed: In order to prevent the residual sign problem, the contribution to the phase coming
from the Jacobian can be included \emph{ab initio}. If the residual phase is small we can follow the previous
principle and ignore it, if it simplifies the expressions.
\item Fixed functional form: The functions used for defining the contour~(\ref{contourForm}) should have the same (relative) functional
dependence on the $x_k$, $\xi_k$, for all the $y_k$, $\zeta_k$. Here, ``relative'' means that the dependence of, say, $y_1$ on, say,
$\xi_2$ would be the same as that of $y_3$ on $\xi_4$.
The motivation behind this restriction in not so much the deep physical principle of homogeneity of space-time
as the ease of implementation. In particular, since the lattice size can vary, we would like to have a prescription for the deformation
that does not change as the lattice size changes.
\item Locality: We would look for contours for which $y_k$, $\zeta_k$ could depend on $x_k$, $\xi_k$ and maybe on their nearest neighbours.
Again, the main motivation is computational rather than the deep physical principle of locality: The evaluation of Jacobians
can become significantly cheaper computationally for local expressions. Also, the parallelisation of local expressions is
much easier and much more efficient.
Moreover, for contours that are defined in terms of some free parameters, locality allows us to fix their values on a small lattice
and use the same values for larger lattices. Of course, the last issues is related to the fact that the original action is indeed
local. 
Physical locality, which in the case of~(\ref{action}) manifests itself as the restriction that interactions are nearest-neighbour ones,
also implies that a contour for which $y_k$, $\zeta_k$ only depend on nearest neighbours could grasp the essential contributions
to the sign problem.
\item Symmetry: The theory we consider here enjoys a $U(1)$ symmetry. It is not clear \emph{a priori} that the best possible contours
should respect this symmetry. After all, the theory has a spontaneously broken phase. However, when examining ansätze for the
form of the contour, one could examine whether the symmetry is respected for ansätze with a small number of parameters.
If this is the case, it could be more efficient to restrict the number of parameters at higher levels in a way that respects the symmetry.
\end{enumerate}
We would like to stress that our approach is practical and that the proposed criteria should be thought of as rules of thumb, to be
implemented only as long as it is beneficial. In the current paper we see that we can benefit from
a mild breaking of translational invariance for our choice of contours\footnote{This does not imply of course that we break
the physical translational invariance, since the integration result does not depend on the choice of the contour and is therefore
not more physical than a gauge choice that breaks a symmetry.}.
This goes against the spirit of points 4 and 6 above, but it turns out that since this breaking is mild the expressions still enjoy
the benefits presented above while another important task (a fast calculation of the Jacobian) is also accomplished.
As another example of breaking these rules of thumb consider a theory with fermions that leads to a bosonic action, which
is not local due to the presence of a fermionic determinant, whose computational cost is high.
It is clear that in such a case one does not have to insist neither on local expressions, since they would not be able to account for
the actual source of the sign problem, nor on a very fast algorithm for evaluating
the Jacobian, since the complexity of the simulation would anyway be restricted by the evaluation of the fermionic determinant.

The rest of the paper is organized as follows:
In section~\ref{sec:simple} we note that the leading source for the sign problem in the current case comes
from nearest neighbour interactions. We then propose to expand the functions~(\ref{contourForm}) with respect to the small parameter
$\alpha$ in order to eliminate this leading contribution to the imaginary part of the action, i.e. we identify
functions~(\ref{contourForm}) that solve~(\ref{imS0}) to lowest order with respect to $\alpha$.
Then, in section~\ref{sec:ansatz}, we generalise these functions to an ansatz that depends on some free parameters.
Next, in section~\ref{sec:Jacobian}, we study the Jacobians related to the contours defined by our ansatz.
Simulation results are presented in section~\ref{sec:Results} and we end with some discussion in section~\ref{sec:conc}.

Note: While this work was prepared the papers~\cite{Mori:2017pne,Mori:2017nwj,Alexandru:2018fqp} appeared,
in which some similar ideas were presented. Our approach shares with~\cite{Alexandru:2018fqp} the low
computational complexity and the ability to use a relatively small number of parameters, while
generalising the ansatz for the contour in a way that enables taking into account the interaction between
nearby lattice points.

\section{Simple integration contours}
\label{sec:simple}

In order to find an approximation for the contour we write the imaginary part of the action~(\ref{S1d})
in terms of the unknown functions $y_k(\{x_m,\xi_m\})$ and $\zeta_k(\{x_m,\xi_m\})$.
If we were to choose
\begin{equation}
\label{contour0}
y_k=\zeta_k=0\,,
\end{equation}
we would have obtained
\begin{equation}
\label{imSnoY}
\im(S)=\frac{\sinh(\mu)}{\lambda\alpha}\sum_{k=1}^L \big(x_k \xi_{k+1}-x_{k+1} \xi_k\big)\,.
\end{equation}
In the following, we refer to the undeformed contour defined by the functions~(\ref{contour0}) as
``contour 0''.

In order to obtain better behaved contours, we expand the imaginary part of the action in powers of $\alpha$.
Recall that the definition of $\alpha$~(\ref{alphaDef}) implies that it is bounded from above by
$\frac{1}{2d}=\frac{1}{2}$ and that the $\alpha\rightarrow 0$ limit, around which we expand,
is the \emph{infinite mass limit}.
We assume that the functions we are after can also be expanded this way and we set their zero order
term to zero, that is we write
\begin{equation}
y_k=\sum_{j=1}^\infty \alpha^j y_k^{(j)}\,.
\end{equation}
In the current paper we will only be interested in the first term in this expansion and in
generalisations of its functional form.
We now write
\begin{equation}
y_k^{(1)}=\sigma\tilde{y}_k,,\qquad
\zeta_k^{(1)}=\sigma\tilde{\zeta}_k\,,
\end{equation}
where we defined
\begin{equation}
\sigma\equiv \alpha \sinh(\mu)\,.
\end{equation}
At the lowest order in the $\alpha$ expansion the imaginary part of the action takes the form,
\begin{align}
\label{ImPart}
\im(S) = \frac{\sinh(\mu)}{\lambda\alpha}\sum_k \bigg( (x_k \xi _{k+1} - x_{k+1} \xi _k) +
\Big(x_k (1+x_k^2+\xi _k^2)
\Big) \tilde{y}_k
 +\Big(\xi_k (1+x_k^2+\xi _k^2)
\Big)\tilde{\zeta }_k\bigg).
\end{align}
Setting this expression to zero gives a single linear equation in the $2L$ unknowns
$\tilde y_k$ and $\tilde \zeta_k$.
Such an equation has many solutions.
However, we are interested only in solutions that are continuous with proper asymptotic behaviour
and for which the dependence of $y_k$ and $\zeta_k$ on $x_k,x_{k+1},\xi_k,\xi_{k+1}$ is the same for all $k$.

We can look for a solution in which the $\tilde y_k$ term cancels the contributions of $x_k \xi _{k+1}$
and the $\tilde\zeta_k$ term cancels the contributions of $- x_{k+1} \xi _k$ for all $k$ even before summation.
Thus, we set,
\begin{subequations}
\label{straightforward}
\begin{align}
\tilde{y}_k &=- \frac{\xi_{k+1}}{1+x_k^2+\xi _k^2}\,,\\
\tilde{\zeta}_k &= \frac{x_{k+1}}{1+x_k^2+\xi _k^2}\,.
\end{align}
\end{subequations}
We refer to the contour defined by~(\ref{straightforward}) as ``contour 1''.
Note that this contour respects the $U(1)$ symmetry of the theory: If we define
\begin{equation}
\phi_k\equiv x_k+i\xi_k	\,, \qquad
\psi_k\equiv y_k+i\zeta_k \,,
\end{equation}
we can write~(\ref{straightforward}) as
\begin{equation}
\label{U1cont1}
\psi_k=\frac{i\sigma\phi_{k+1}}{1+|\phi_k|^2}\,.
\end{equation}
The $U(1)$ symmetry rotates both $\phi_k$ and $\psi_k$ in the standard manner.

Nonetheless, contour 1 does break a symmetry.
Sending $k+1\rightarrow k-1$ while either taking a complex conjugation or sending $\mu\rightarrow -\mu$,
leaves the action~(\ref{S1d}) invariant.
This discrete symmetry is broken by the contour. One could act with this symmetry on the contour
obtaining an equivalent one defined by,
\begin{equation}
\psi_k=-\frac{i\sigma\phi_{k-1}}{1+|\phi_k|^2}\,.
\end{equation}
This contour could have also been obtained by a rewriting of the terms in the
sum before deciding that they should vanish even before summation.
However, since this contour is equivalent to contour 1, we do not study it separately.

The expressions~(\ref{straightforward}) are not bounded in the limits $\xi_{k+1}\rightarrow\infty$ and
$x_{k+1}\rightarrow\infty$ respectively, with all other variables fixed.
This means that the asymptotic range covered by them might include also parts of the complexified space
unrelated to the original contour. Nonetheless, since these expressions can be obtained by continuously deforming
the original integration contour, they would lead to the correct results, as long as they do not pass
asymptotic regions over which the integral diverges.
One could further worry that even if the integrals along these parts of the contour are finite,
they would include large contributions that are supposed to cancel each other.
The main contribution to the different phase in this case would come from the Jacobian.
Such a scenario could lead to a
global sign problem reminiscent of problems with integration along Lefschetz thimbles.
For similar reasons integration along a generalised thimble can behave better than integration along
the thimble itself~\cite{Alexandru:2018ddf,Lawrence:2018mve}.
This problem could be addressed by slightly modifying the contour in order to obtain
expressions that are everywhere bounded. A simple possibility is to use the following expressions,
\begin{subequations}
\label{yZetak0}
\begin{align}
\tilde{y}_k &=- \frac{\xi_{k+1}}{1+x_k^2+\xi_k^2+\xi_{k+1}^2}\,,\\
\tilde{\zeta}_k &= \frac{x_{k+1}}{1+x_k^2+\xi_k^2+x_{k+1}^2}\,.
\end{align}
\end{subequations}
These expressions neither respect the $U(1)$ symmetry
nor do they lead to $\im(S)=0$ even at the lowest order of our expansion.
Instead, they respect a $\Z_4$ subgroup of $U(1)$, generated by
\begin{equation}
\label{Z4}
x_k\rightarrow \xi_k\,,\qquad \xi_k\rightarrow - x_k\,,
\end{equation}
while the imaginary part of the action is given by,
\begin{align}
\im(S)=\frac{\sinh(\mu)}{\lambda\alpha} \sum_k \bigg(\frac{x_k \xi_{k+1}^3
}{1+x_k^2+\xi_k^2+\xi_{k+1}^2} 
  - \frac{\xi_k x_{k+1}^3
}{1+x_k^2+\xi_k^2+x_{k+1}^2}\bigg).
\end{align}
However, it might happen that the benefit from avoiding infinity is more significant for taming the sign problem than this small
mismatch. We refer to the contour defined by~(\ref{yZetak0}) as ``contour 2''.

Of course, in order to turn~(\ref{straightforward}) into bounded expressions, one could also use variants of~(\ref{yZetak0}).
In particular, one could extrapolate between~(\ref{straightforward}) and~(\ref{yZetak0}) by introducing a constant $\kappa$
and setting
\begin{subequations}
\label{yZetak3}
\begin{align}
\tilde{y}_k &=- \frac{\xi_{k+1}}{1+x_k^2+\xi_k^2+\kappa\xi_{k+1}^2}\,,\\
\tilde{\zeta}_k &= \frac{x_{k+1}}{1+x_k^2+\xi_k^2+\kappa x_{k+1}^2}\,.
\end{align}
\end{subequations}
Now, for $\kappa=0$ we obtain~(\ref{straightforward}), while~(\ref{yZetak0}) is obtained for $\kappa=1$
and values of $\kappa>1$ can also be considered.
We could try to look for the value of $\kappa$ for which the sign problem is most significantly reduced.
Eq.~(\ref{yZetak3}) is an ansatz for the contour depending on a single parameter.
We refer to this 1-parameter ansatz with an optimal choice of $\kappa$ as ``contour 3''.
We now turn to define a more general ansatz.

\section{A general ansatz for the integration contour}
\label{sec:ansatz}

In order to obtain better behaved contours,
it might be useful to generalise the expressions considered so far.
A generalised contour could account for approximations we have made, as well as for the phase factor
due to the Jacobian, that we did not considered in the discussion so far.
A natural generalisation would be to rely on the functional forms defining the contours proposed above, while changing the powers of the
various factors and allowing for a somewhat less local functional dependence.
One can see that if we were to go to higher orders in our $\alpha$ expansion, such terms would have indeed appeared.
We therefore consider the following functional form for defining the
contour\footnote{Note that $\mu$, $\nu$, $\rho$, $\sigma$, $\tau$, $\phi$ are powers to which the variables are raised,
not indices, which would have made no sense in the current one-dimensional case.},
\begin{subequations}
\label{generalcontour}
\begin{align}
\tilde{y}_k &= - \frac{\sum_{\mu \nu \rho \sigma \tau \phi} a_{\mu \nu \rho \sigma \tau \phi} x_k^\mu x_{k-1}^\nu x_{k+1}^\rho \xi_k^\sigma \xi_{k-1}^\tau \xi_{k+1}^\phi}
    {1 + \sum_{\mu \nu \rho \sigma \tau \phi} b_{\mu \nu \rho \sigma \tau \phi} x_k^\mu x_{k-1}^\nu x_{k+1}^\rho \xi_k^\sigma \xi_{k-1}^\tau \xi_{k+1}^\phi}\,, \\
\tilde{\zeta}_k &=
\frac{\sum_{\mu \nu \rho \sigma \tau \phi} a_{\mu \nu \rho \sigma \tau \phi} x_k^\sigma x_{k-1}^\tau x_{k+1}^\phi \xi_k^\mu \xi_{k-1}^\nu \xi_{k+1}^\rho}
    {1 + \sum_{\mu \nu \rho \sigma \tau \phi} b_{\mu \nu \rho \sigma \tau \phi} x_k^\sigma x_{k-1}^\tau x_{k+1}^\phi \xi_k^\mu \xi_{k-1}^\nu \xi_{k+1}^\rho}\,.
\end{align}
\end{subequations}
Here the sum in the numerator runs over all non-negative integers
$\mu, \nu, \rho, \sigma, \tau, \phi$ such that $0 \le \mu + \nu + \rho + \sigma + \tau + \phi \le p$,
and the sum in denominator runs over all \emph{even} non-negative integers
$\mu, \nu, \rho, \sigma, \tau, \phi$ such that $2 \le \mu + \nu + \rho + \sigma + \tau + \phi \le q$.
We define $(p, q)$ to be the order of the contour.
The reason for restricting the sum to only even integers in the denominator is to ensure that
the denominator is always positive so that the contour always remains continuous for finite values of the integration variable.
This also requires that the $b_{\mu \nu \rho \sigma \tau \phi}$ are non-negative.
The requirement of obtaining proper boundary conditions implies the inequality $q\geq p$.
One can further restrict the ansatz by requiring that no terms
appear in the numerator with powers of any variable that does not appear in the denominator
with at least the same power. This will result in contours that are everywhere bounded.

The ansatz~(\ref{generalcontour}) includes all contours considered so far and generalises them using rational functions.
One advantage of using rational function as in~(\ref{generalcontour}) is that they are easy to differentiate and so the Jacobian
can be explicitly written.
The ansatz also has a sign choice that reminds of the previously considered contours. Specifically, if one only considers non-zero coefficients
in the numerator for even values of $\mu+\nu+\rho$ and odd values of $\sigma+\tau+\phi$, the $\Z_4$ symmetry~(\ref{Z4}) is respected.
A further possible restriction is to impose the whole $U(1)$ symmetry,
that is to consider only expressions that can be written in a way that generalises~(\ref{U1cont1}).
Such a restriction would not necessarily lead to the best contours.
However, if it turns out that the $U(1)$ symmetry gives approximately the best contours at low contour order, it would make sense to
simplify the optimisation procedure by imposing the $U(1)$ symmetry on the ansatz
\emph{ab initio}, reducing in this way the amount of free parameters needed to be optimised.

It is easy to include higher order terms in this contour simply by increasing the order.
Similarly, longer-range terms could be included by adding to the ansatz expressions depending on $x_{k \pm 2}$, $\xi_{k \pm 2}$ etc.
However, this comes at a cost; we need to determine the parameters $a_{\mu \nu \rho \sigma \tau \phi}, b_{\mu \nu \rho \sigma \tau \phi}$.
There may be a large number of these; without imposing restrictions on the coefficients,
already at order $(2, 2)$ there are 34 free parameters.
Furthermore, the optimal contour will presumably depend on the chemical potential $\mu$, the interaction strength $\lambda$, and on $\alpha$.

Another important restriction is to contours for which the Jacobian can be efficiently evaluated. We discuss this issue in the next section.
For now we just mention that this restriction implies that one should consider in~(\ref{generalcontour}) either only terms with $x_{k+1}$ and $\xi_{k+1}$
or terms with $x_{k-1}$ and $\xi_{k-1}$, but not both. We refer in what follows to the use of only $x_{k+1}$ and $\xi_{k+1}$ as
``forward nearest neighbour''
and to a forward nearest neighbour ansatz with optimised parameters as ``contour 4''. An optimised choice for ansatz parameters without this restriction
would be referred to as ``contour 5''.

Choosing the best contour sounds like a daunting task; on a large lattice where the sign problem is severe we would expect that for generic contour
parameters $a_{\mu \nu \rho \sigma \tau \phi}, b_{\mu \nu \rho \sigma \tau \phi}$ the mean phase factor will be so small that it will be statistically indistinguishable from zero.
How can we then decide how to change the parameters to obtain a larger mean phase factor?
The solution is to tune the contour on small lattices. Since the action is local, the imaginary part of the action is presumably mostly due to
local contributions rather than long-range correlations. Our ansatz is also local. Hence, if we tune the parameters
$a_{\mu \nu \rho \sigma \tau \phi}, b_{\mu \nu \rho \sigma \tau \phi}$ to maximise the mean phase factor
on a small volume we will obtain a contour that minimises these contributions.
Then, if we use a contour defined by the same parameters
on a larger volume the imaginary part of the action should still be small.
If needed, one could start on a large lattice with the parameters fixed on a small one and update them, but
locality implies that generically this would not lead to a significant improvement.
We explain the methods used for optimising the ansatz parameters in subsection~\ref{sec:fit}.

\section{The Jacobian}
\label{sec:Jacobian}

To do calculations with these contours we must parametrise $u_k$ and $v_k$ in some convenient way.
The most natural thing to do is to use the real parts $x_k$ and $\xi_k$.
We must then include the Jacobian determinant in the evaluation,
\begin{equation}
\mathrm{det} (J) = \mathrm{det} \left( \begin{array}{ccccc} 
\frac{\partial u_1}{\partial x_1} & \frac{\partial u_1}{\partial \xi_1} & \frac{\partial u_1}{\partial x_2} & \frac{\partial u_1}{\partial \xi_2} & \dots\\
 \frac{\partial v_1}{\partial x_1} & \frac{\partial v_1}{\partial \xi_1} & \frac{\partial v_1}{\partial x_2} & \frac{\partial v_1}{\partial \xi_2} & \dots\\
\frac{\partial u_2}{\partial x_1} & \frac{\partial u_2}{\partial \xi_1} & \frac{\partial u_2}{\partial x_2} & \frac{\partial u_2}{\partial \xi_2} & \dots \\
 \frac{\partial v_2}{\partial x_1} & \frac{\partial v_2}{\partial \xi_1} & \frac{\partial v_2}{\partial x_2} & \frac{\partial v_2}{\partial \xi_2} & \dots\\
\vdots & \vdots & \vdots & \vdots & \ddots \\
\end{array} \right)
\end{equation}
For contour 1~(\ref{straightforward}) the Jacobian matrix takes the form,
\begin{equation}
\label{Jcont1}
J = \left( \begin{array}{cccccccc} 
1 + 2 i \sigma \frac{x_1 \xi_2}{d_1^2} & 2 i \sigma \frac{\xi_1 \xi_2}{d_1^2} & 0 & -i \sigma \frac{1}{d_1} & 0 & 0 & 0 & \dots\\
-2 i \sigma \frac{x_1 x_2}{d_1^2} & 1 - 2 i \sigma \frac{\xi_1 x_2}{d_1^2} & i \sigma \frac{1}{d_1} & 0 & 0 & 0 & 0 & \dots\\
0 & 0 & 1 + 2 i \sigma \frac{x_2 \xi_3}{d_2^2} & 2 i \sigma \frac{\xi_2 \xi_3}{d_2^2} & 0 & -i \sigma \frac{1}{d_2} & 0 & \dots \\
0 & 0 & -2 i \sigma \frac{x_2 x_3}{d_2^2} & 1 - 2 i \sigma \frac{\xi_2 x_3}{d_2^2} & i \sigma \frac{1}{d_2} & 0 & 0 & \dots\\
\vdots & \vdots & & & \ddots & \ddots & \\
0 & -i \sigma \frac{1}{d_L} & 0 & 0 & \dots & & & \\
i \sigma \frac{1}{d_L} & 0 & 0 & 0 & \dots & & & \\
\end{array} \right)
\end{equation}
where we defined
\begin{equation}
d_k \equiv 1+x_k^2+\xi_k^2\,.
\end{equation}
For contour 2~(\ref{yZetak0}) we have a slightly more complicated Jacobian matrix,
\begin{align}
\label{Jcont2}
J= \!\! \left( \! \begin{array}{cccccccc} 
1 + 2 i \sigma \frac{x_1 \xi_2}{\tilde{d}_1^2} & 2 i \sigma \frac{\xi_1 \xi_2}{\tilde{d}_1^2} & 0 & -i \sigma \frac{\tilde{d}_1-2\xi_2^2}{\tilde{d}_1^2} & 0 & 0 & 0 & \dots\\
-2 i \sigma \frac{x_1 x_2}{\hat{d}_1^2} & 1 - 2 i \sigma \frac{\xi_1 x_2}{\hat{d}_1^2} & i \sigma \frac{\hat{d}_1-2x_2^2}{\hat{d}_1^2} & 0 & 0 & 0 & 0 & \dots\\
0 & 0 & 1 + 2 i \sigma \frac{x_2 \xi_3}{\tilde{d}_2^2} & 2 i \sigma \frac{\xi_2 \xi_3}{\tilde{d}_2^2} & 0 & -i \sigma \frac{\tilde{d}_2-2\xi_3^2}{\tilde{d}_2^2} & 0 & \dots \\
0 & 0 & -2 i \sigma \frac{x_2 x_3}{\hat{d}_2^2} & 1 - 2 i \sigma \frac{\xi_2 x_3}{\hat{d}_2^2} & i \sigma \frac{\hat{d}_2-2x_3^2}{\hat{d}_2^2} & 0 & 0 & \dots\\
\vdots & \vdots & & & \ddots & \ddots & \\
0 & -i \sigma \frac{\tilde{d}_L-2\xi_1^2}{\tilde{d}_L^2} & 0 & 0 & \dots & & & \\
i \sigma \frac{\hat{d}_L-2x_1^2}{\hat{d}_L^2} & 0 & 0 & 0 & \dots & & & \\
\end{array} \! \right)
\end{align}
where now we defined
\begin{equation}
\tilde{d}_k  = 1+x_k^2+\xi_k^2 + \xi_{k+1}^2\,,\qquad \hat{d}_k = 1+x_k^2+\xi_k^2 + x_{k+1}^2\,.
\end{equation}
Similarly it is straightforward to derive the Jacobian for our general ansatz (\ref{generalcontour}).

We note, that both~(\ref{Jcont1}) and~(\ref{Jcont2}) are almost upper-block-bidiagonal with $2\times 2$ blocks.
The block structure comes from the fact that we have two fields in every lattice site and the upper-bidiagonal structure
originates from the use of forward nearest neighbour in the definition of the contours.
Thus, in the general forward nearest neighbour case, the Jacobian matrix takes the form
\begin{equation}
J= \left( \begin{array}{ccccc} 
A_1 & B_1 & 0 & 0 & \dots\\
0 & A_2 & B_2 & 0 & \dots\\
\vdots & \ddots & \ddots & \ddots & 0\\
0 & \dots & 0 & A_{L-1} & B_{L-1}\\
\red{B_L} & 0 & \dots & 0 & A_L
\end{array} \right)
\end{equation}
The only term that does not fit the upper-bidiagonal form is
$B_L$ (marked in red).
This term is there in light of the periodic boundary conditions.
Had  this term been absent, the Jacobian determinant would be the product of $2\times 2$ determinants.
However, it is easy to evaluate the Jacobian determinant even when this term is present.
If the blocks were just numbers we could
have used elementary row operations in order to eliminate $B_{L-1}$, then eliminate $B_{L-2}$, and so on, obtaining a matrix
whose determinant is given by
\begin{equation}
\det(J)= \left(A_1 - (-1)^L B_1 A_2^{-1}\cdot ...\cdot B_{L-1}A_L^{-1}B_L\right)\cdot A_2\cdot...\cdot A_L\,.
\end{equation}
In the current case where we deal with blocks instead of numbers, we can perform ``elementary row operations'' by left multiplication by
``elementary block matrices'', whose determinant is unity.
This leads to a very similar expression,
\begin{equation}
\begin{aligned}
\det(J) &= \det\left(A_1 - (-1)^L B_1 A_2^{-1}\cdot ...\cdot B_{L-1}A_L^{-1}B_L\right)\cdot \det\left(A_2\right)\cdot...\cdot \det\left(A_L\right)\\
        &= \det\left(\id - (-1)^L S_1 S_2 \cdot ...\cdot S_L\right)\cdot \det\left(A_1\right)\cdot...\cdot \det\left(A_L\right)\,,
\end{aligned}
\end{equation}
where we defined
\begin{equation}
S_k \equiv A_k^{-1}B_k\,.
\label{Sk}
\end{equation}

We reduced the evaluation of the determinant to the evaluation of a product of block matrices.
Thus we resolved the complexity related to the evaluation of the determinant.
One could still worry that since the determinant includes a product of $L+1$ terms of which the evaluation of the first involves a product of $L$ matrices
it would lead to a complexity of $O(L^2)$.
This is not the case. The locality of our prescription implies that when we change $x_k$ or $\xi_k$
we should only update $A_{k-1}$, $A_k$, $B_{k-1}$, $B_k$. Dividing the product by the previous value of $\det\left(A_{k-1}\right)\det\left(A_k\right)$
and multiplying by the new value of this factor is of $O(1)$. As for the first term in the product, symmetry of the problem implies that it can be written in any cyclic order,
\begin{equation}
\label{CyclicM}
\det\left(\id - (-1)^L S_1 S_2\cdot ...\cdot S_{L-1}S_L\right)=
\det\left(\id - (-1)^L S_k S_{k+1}\cdot ...\cdot S_{k-2}S_{k-1}\right)\,.
\end{equation}
This identity can also be proven explicitly, for example by expressing the determinant in terms of a trace and utilizing the cyclicity property of the trace.
Thus, when we initiate the simulation we define
\begin{equation}
\label{Mdef}
M\equiv S_L S_1 \cdot ...\cdot S_{L-1}\,.
\end{equation}
We start by updating $x_1$ after which we change
\begin{equation}
M\rightarrow \big(S_L S_1\big)_{\new}\big(S_L S_1\big)_{\old}^{-1} M\,.
\end{equation}
We then update $\xi_1$ after which we change\footnote{In fact, we update $x_k$ and $\xi_k$ at a single step, so we only have to use
the second update. Note, that one must perform the second update even when the update for $x_k$, $\xi_k$ is rejected, since
it brings the matrix $M$ to the form needed for updating the next site.},
\begin{equation}
\label{newM}
M\rightarrow \big(S_1\big)_{\new}\big(S_L S_1\big)_{\old}^{-1} M\big(S_L\big)_{\new}\,.
\end{equation}
We used the cyclicity property~(\ref{CyclicM}) in order to obtain an expression that is adequate for $x_2$.
We then continue with $x_2$, $\xi_2$ and so on. All the manipulations described are of $O(1)$ and so do not increase the complexity.

The above algorithm might not always work: While the $A_k$ are always invertible the $B_k$ can be non-invertible for specific contours.
Moreover, even when the $B_k$ are invertible, they might be close to being singular. This, together with the fact that these are complex matrices,
could lead to stability issues for the algorithm.
There are several options for handling this obstacle:
\begin{enumerate}
\item One could use a slow algorithm to evaluate the determinant without using the nearly-bidiagonal structure,
such as LU decomposition ($O(L^3)$ per site so $O(L^4)$ per sweep),
or evaluate $M$ using its definition~(\ref{Mdef}) after every update, which leads to complexity of $O(L^2)$.
This solution is, of course, not the desired one, since even the $O(L^2)$ would be much more restrictive than an $O(L)$ algorithm.
\item The problem described would emerge only for specific values of the parameters defining the general contour.
One could then choose a different set of parameters, for which the $B_k$ are generically large and the problem does not occur.
\item It is possible to deform all the contours, but one. Suppose we decide not to deform $u_1,v_1$, this modifies the expression for the
Jacobian to the simpler form,
\begin{equation}
\label{secondApproachJ}
\det(J) = \det\left(A_2\right)\cdot...\cdot \det\left(A_L\right)\,.\\
\end{equation}
While such a choice would certainly decrease the mean phase factor, we expect that this change would be $L$-independent and so would not change by much the
range of validity of the given contour, i.e. the unmodified and modified contours should begin to suffer from the sign problem at not too
different values of $L$.
If this option is chosen one should verify that observables at $k=1$ do not behave in any
anomalous way. While analytically this is obvious, the asymmetric treatment of a single point can in principle
induce different numerical errors in this point as compared to the other ones.
\item Since the numerical instability would usually occur for small values of $B_k$, one could expect that for large enough $L$,
the matrix $M$ would become extremely small and the approximation
\begin{equation}
\label{approxM}
\det(J)\simeq \det\left(A_1\right)\cdot...\cdot \det\left(A_L\right) 
\end{equation}
would become exact within the simulation precision.
This expression is almost identical to~(\ref{secondApproachJ}), but the decrease of the phase factor by a constant is now absent.
Moreover, since in this case (as well as in the previous one) there is no need to evaluate the matrix $M$, the computational
cost is further reduced.
\end{enumerate}
All these options are examined in subsection~\ref{sec:resultsInstability}.

The discussion so far is inadequate for the general form for the contour~(\ref{generalcontour}),
without imposing the forward nearest neighbour restriction.
In this general case the Jacobian matrix is almost tridiagonal-block-matrix and we have no fast algorithm for evaluating it.
In the next section we examine the various contours, including contour 5, for which we use a slow algorithm.
It turns out that there is no significant improvement in going from contour 4 to contour 5.
The amount of free parameters in contour 4 is much smaller than in contour 5.
Hence, the restriction to forward nearest neighbour contours could even enable one to get to a better resolution of the sign
problem given a fixed number of free parameters, since then higher $(p,q)$ values can be considered.

\section{Results}
\label{sec:Results}

As mentioned in the introduction, we examine the method presented in this paper by simulating the system
with $\lambda=m=\mu=1$.
The simulation is performed using a Fortran code.
In all simulations we take 300,000 configurations. The jackknife method is used for the evaluation of statistical errors.
We examine the observables $\vev{S}$ and $\vev{u_k^2+v_k^2}$ as a function of lattice size for the different contours suggested.
Since we only illustrate the method we do not study other observables. Of course, one could examine in this way
also the average density and other observables.
The mean phase factor for the same contours is also studied, since it serves as an indicator to the severity of the sign problem.
As expected, this factor decreases as the lattice size is increased.
We find high correlation between problems with the observables we examine and small values of the mean phase factor.
Thus, it is natural to expect that problems related to the sign problem with other observables would begin to emerge
roughly for the same lattice size as for the observables studied here.

We simulate all the contours described above: the undeformed contour 0~(\ref{contour0}),
contour 1~(\ref{straightforward}), contour 2~(\ref{yZetak0}), contour 3~(\ref{yZetak3}), contour 4, which is
a forward nearest neighbour version of the general ansatz~(\ref{generalcontour}) for $(p,q)=(1,2)$, and contour 5, which is
an unrestricted version of the general ansatz~(\ref{generalcontour}) for $(p,q)=(1,2)$.
We use the fast algorithm for the evaluation of the Jacobian presented in section~\ref{sec:Jacobian} for all contours,
except contour 5, for which it is not applicable.

Contours 3,4,5 are representatives within some given ansätze.
In subsection~\ref{sec:fit}, we present the approach we use for deciding which values should the parameters take.
Next, in subsection~\ref{sec:resultsMain}, we present the main results of our simulations. First,
we examine all the contours for $L=4,8,16,24,32,...,96$. Then, we offer a prediction for the
onset of the sign problem for the various contours, which we examine using $L=100,150,200,250,...,1000$.
Since contour 5 is evaluated using a slow algorithm, we do not examine this prediction for this contour.
As mentioned in section~\ref{sec:Jacobian}, the fast algorithm for evaluating contours 1,2,3,4 may suffer
from numerical instability. We monitor this issue in our simulations.
To that end, we evaluated $M$ using its original definition~(\ref{Mdef}) at the end of each sweep
and compared to the result obtained by successively using the update algorithm~(\ref{newM})\footnote{Note, that since this check
is performed at the end of a sweep, it has a computational cost of $O(L)$ so it does not increase the asymptotic cost.}.
While contours 1,2,3 show no numerical instability, for contour 4 the relative size of the components of $M$,
evaluated in these two ways, differs from unity by more the one part per million
(the threshold we used), already at $L=16$. We examine possible resolutions of the numerical instability
in subsection~\ref{sec:resultsInstability}.

\subsection{Fitting the parameters}
\label{sec:fit}

An important characteristic of our approach is the locality of the contours and of the action,
which implies that values of the parameters that minimize the sign problem for some large enough $L$, would
also minimize the sign problem for higher values of $L$. This fact enables us to identify the parameters
for a given value of $L$, for which the evaluation is fast, and use the same values for all values of $L$.
We then verify that small changes of the parameters for larger values of $L$ do not lead to further improvement
with dealing with the sign problem.
In all these evaluations, we use the mean phase factor as a measure of the severity of the sign problem.

Finding the best fit for defining contour 3 is quite simple, since there is only a single parameter in this case.
We examine $L=16$ for different values of $\kappa$ and find that the mean phase factor is roughly constant in the range $0<\kappa<0.5$
taking values in the range $(0.708-0.715)$
and begins to drop as $\kappa$ is further increased. We choose the middle of this range, $\kappa=0.25$ as the representative
for defining contour 3, although slightly larger values for the mean phase factor were obtained both above and below
this value. However, the difference is very mild and could well be of statistical nature.

Contours 4 and 5 have several parameters and finding them is less trivial.
Again, we use the locality for finding the values of the parameters for a fixed $L$, this time for $L=8$.
To that end, optimisation of the contour was performed one parameter at a time,
by simulating for several values of the parameter, fitting a quadratic to the mean phases obtained,
and updating the parameter to the maximum of this quadratic.
More efficient methods, e.g. steepest ascent, could also be used in principle.
However, we find that convergence using this method is very rapid,
typically within 10 or so iterations, so using more efficient methods is unnecessary.
Using this method, we obtain the following values for the parameters defining contour 4,
\begin{equation}
\begin{aligned}
\label{5f}
& a_{000001} = 0.56\,,\qquad & a_{000100} = 1.88\,,\qquad & \\
& b_{000002} = 0.03\,,\qquad & b_{000200} = 0.20\,,\qquad & b_{002000} = 0.05\,,\qquad & b_{200000} = 0.19\,,
\end{aligned}
\end{equation}
with all other contour parameters equal to zero.
Notice that the parameters are very close to a $U(1)$ symmetric contour, i.e. $b_{000002}\simeq b_{002000}$
and $b_{000200}\simeq b_{200000}$.
We use this fact in subsection~\ref{sec:resultsInstability} where we look for alternative values for the
parameters. There, we preform a search in the space of $U(1)$ symmetric contours, which is defined by only four free parameters.
In this search we use a Metropolis-like approach: We choose a point in the space of the four
free parameters and run the simulation with these values. We then consider a nearby configuration in this space
and decide whether to move to this point or not using the weight function $\frac{1}{1-\vev{e^{i\im(S)}}_{PQ}}$.
Other functions that become large near the optimal mean phase value of unity could also be used as weight functions
and again, other methods could also be used.
However, this is not needed, since results are easily obtained with the current approach.
During this random walk we keep track of values of the parameters that lead to mean phase
factors higher than some chosen cutoff value.

For contour 5 we use again the initial approach used for defining contour 4 and find,
\begin{equation}
\begin{aligned}
& a_{000001} = 0.73\,,\qquad & a_{000010} = 0.15\,,\qquad & a_{000100} = 1.77\,,\\
& b_{000020} = 0.03\,,\qquad & b_{000200} = 0.27\,,\qquad & b_{002000} = 0.05\,,\qquad & b_{200000} = 0.28\,,
\end{aligned}
\end{equation}
with all other contour parameters equal to zero.
We see that these values are not too far from the ones of contour 4 and the values that are inconsistent with
a forward nearest neighbour contour are small. Thus, we expect that the benefit from using the slow contour 5 over
contour 4 is not too high.

One could wonder about the values of the parameters obtained: They define contours that differ quite a lot from
contours 1,2,3. In particular the weight of $x_k$ and $\xi_k$ in the numerators is higher than that of $x_{k+1}$ and $\xi_{k+1}$.
What could be the reason for that?
We propose the following observation: An exact and simple solution to the equation $\im(S)=0$ exists, which also
leads to a constant Jacobian and hence could seem as a complete resolution of the sign problem,
\begin{equation}
y_k=-\xi_k\,,\qquad \zeta_k=x_k\,.
\end{equation}
However, not only $\im(S)=0$ on this contour, but actually $S=0$. Hence, it neither leads to a convergent
action nor does it obey the proper boundary conditions and is therefore inadequate as an integration contour.
Nonetheless, it might be beneficial to go in the direction of such a contour before changing the course
towards the one of contour 1. It seems to us that contours 4 and 5 give such an interpolation between these two contours.

\subsection{The observables and the mean phase factor for the various contours}
\label{sec:resultsMain}

\begin{figure}[ht]
\includegraphics[width=5.95in]{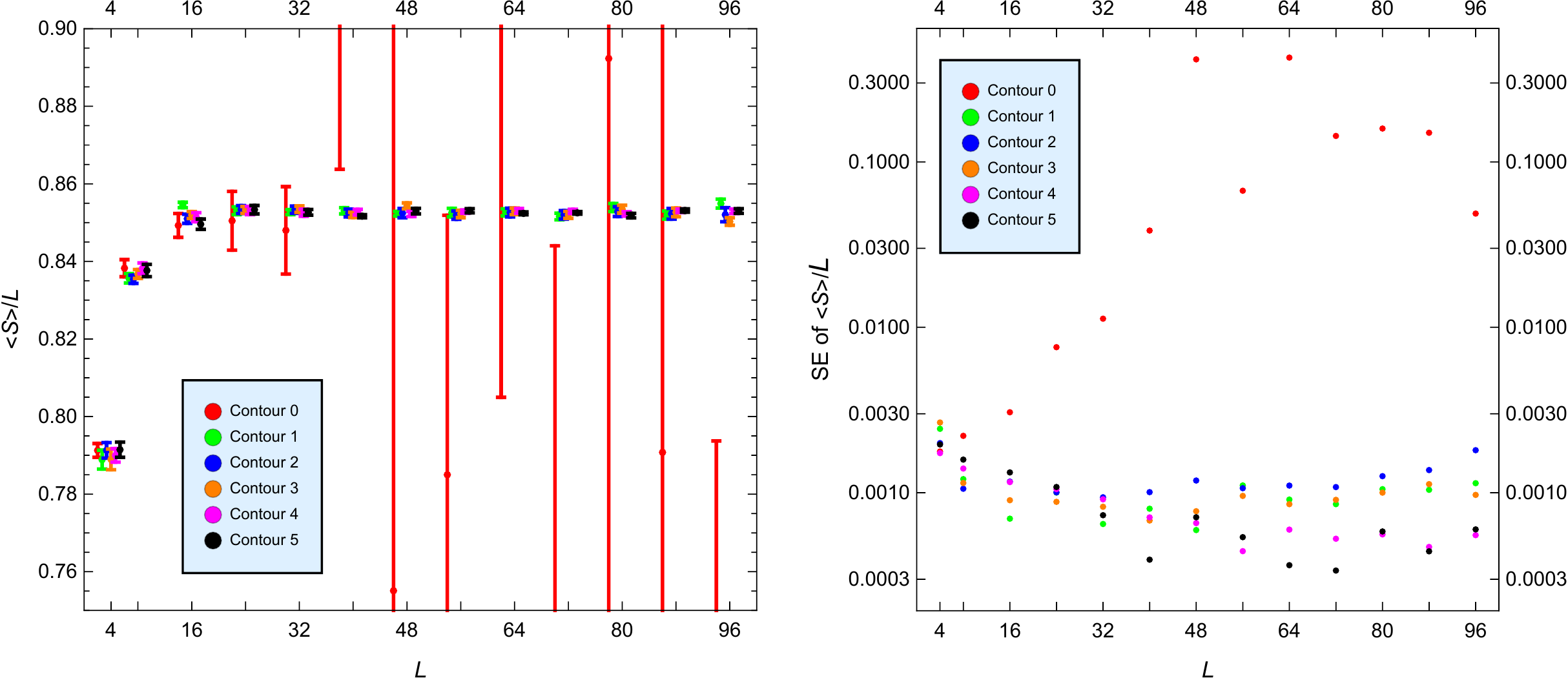}
\caption{Left: The action density as a function of lattice size.
Different contours are separated horizontally for clarity.
Right: The standard error of the action density.
The various contours are colour marked.}
\label{fig:action}
\end{figure}

In fig.~\ref{fig:action} we plot the action density, that is, the expectation value of the action divided by the lattice size,
for the various contours described above.
The different contours are horizontally separated to enable distinguishing them, as many of them overlap.
Since it is hard to infer the standard errors from the plot they are also displayed separately.
It is obvious that one cannot trust contour 0 for $L\gtrsim 32$.
As for the other contours, they all give the same results, which are consistent with the expectation
of obtaining a constant action density for large $L$.
Given the deformation of the contours and the fact that the action is complex, we could have obtained general
complex results. In fact, all the results are consistent with being real. The largest value for $\frac{\left|\im\vev{S}\right|}{L}$
is 0.1 for contour 0 and about 0.004 for contour 3. It is smaller for the other contours.
This gives further support to our method.

It seems that there is no significant difference between contours 1,2,3 as far as the standard error
of the action is concerned. Initially it goes down, but then, around $L=40$, it starts to go up again.
The situation with contours 4 and 5 is less clear, since the tendencies are masked by noise.
It is hard to infer whether they started going up already or not.
In general, there can be several reasons for a dependence of the standard error on lattice size:
Since we are not at a critical point, observables evaluated at sites separated by more than a finite correlation length are roughly independent.
We then expect the standard error of the action to scale like that of $L/l_{\mathrm{corr}}$ independent measurements of the same object, that is, like $\sqrt{L}$.
Thus, if this was the only relevant effect, we would have expected the standard error of the action density to scale
like $L^{-\frac{1}{2}}$. However, there are two effects that increase the size of the standard error:
the increase of the autocorrelation time
and the sign problem. Thus, one could suspect that contours
1,2,3 are closer to developing a sign problem than contours 4 and 5, although
it is hard to make this statement into a quantitative
one based solely on the values of the standard error.

\begin{figure}[ht]
\includegraphics[width=5.95in]{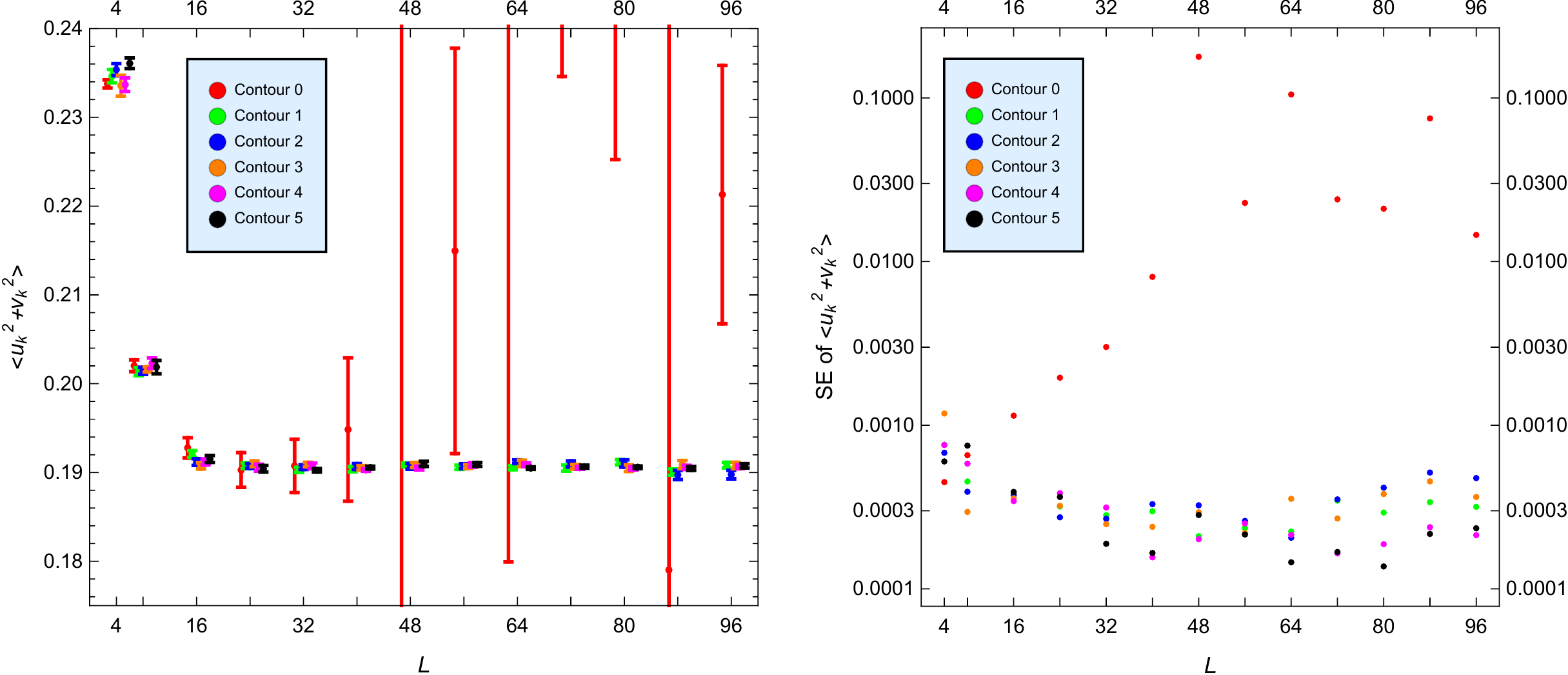}
\caption{Left: $\vev{u_k^2+v_k^2}$ as a function of lattice size.
Different contours are separated horizontally for clarity.
Right: The standard error of $\vev{u_k^2+v_k^2}$. The various contours are colour marked.}
\label{fig:sumsq}
\end{figure}

In fig.~\ref{fig:sumsq} we plot the expectation value of $u_k^2+v_k^2$ as a function of lattice size.
The general tendencies identified for the action can be seen also here.
The results of contour 0 are slightly better now, presumably since no higher order terms are involved.

\begin{figure}[ht]
\begin{center}
  \includegraphics[width=4.7in]{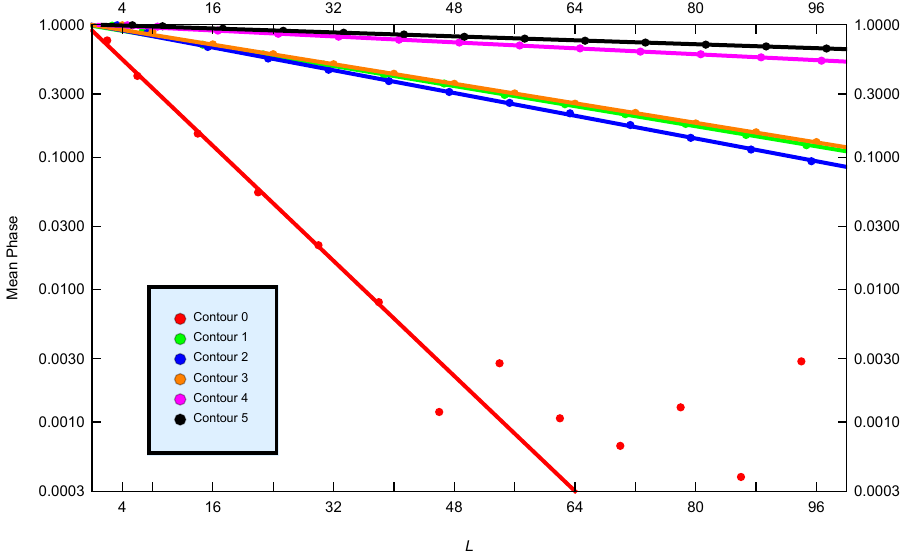}
\caption{The mean phase factor as a function of lattice size on a logarithmic scale for the different contours
together with their linear least square fits.
The various contours are colour marked. Different contours are separated horizontally for clarity.}
\label{fig:phase}
\end{center}
\vspace{-2mm}
\end{figure}

In fig.~\ref{fig:phase} we plot the mean phase factor as a function of lattice size.
One can see that on this logarithmic scale it decreases linearly for all contours
(the lines plotted are mean square fits of the data),
until, for contour 0, it gets to the range, in which we already found out that the sign problem kicks in.
This loss of linear behaviour (on a logarithmic scale) that signifies the sign problem happens around a mean phase of about 0.01.
It is natural to anticipate that, at least for a given theory, similar values of the mean phase factor would be associated with the sign problem
for different integration contours.
Thus, by extrapolating the plotted lines to 0.01 one can attempt to predict the onset
of the sign problem for the other contours.
We therefore expect that the onset of the sign problem would occur around $L=200$ for contour 1,2,3,
around $L=700$ for contour 4, and around $L=1100$ for contour 5.

\begin{figure}[ht]
\begin{center}
  \includegraphics[width=5.95in]{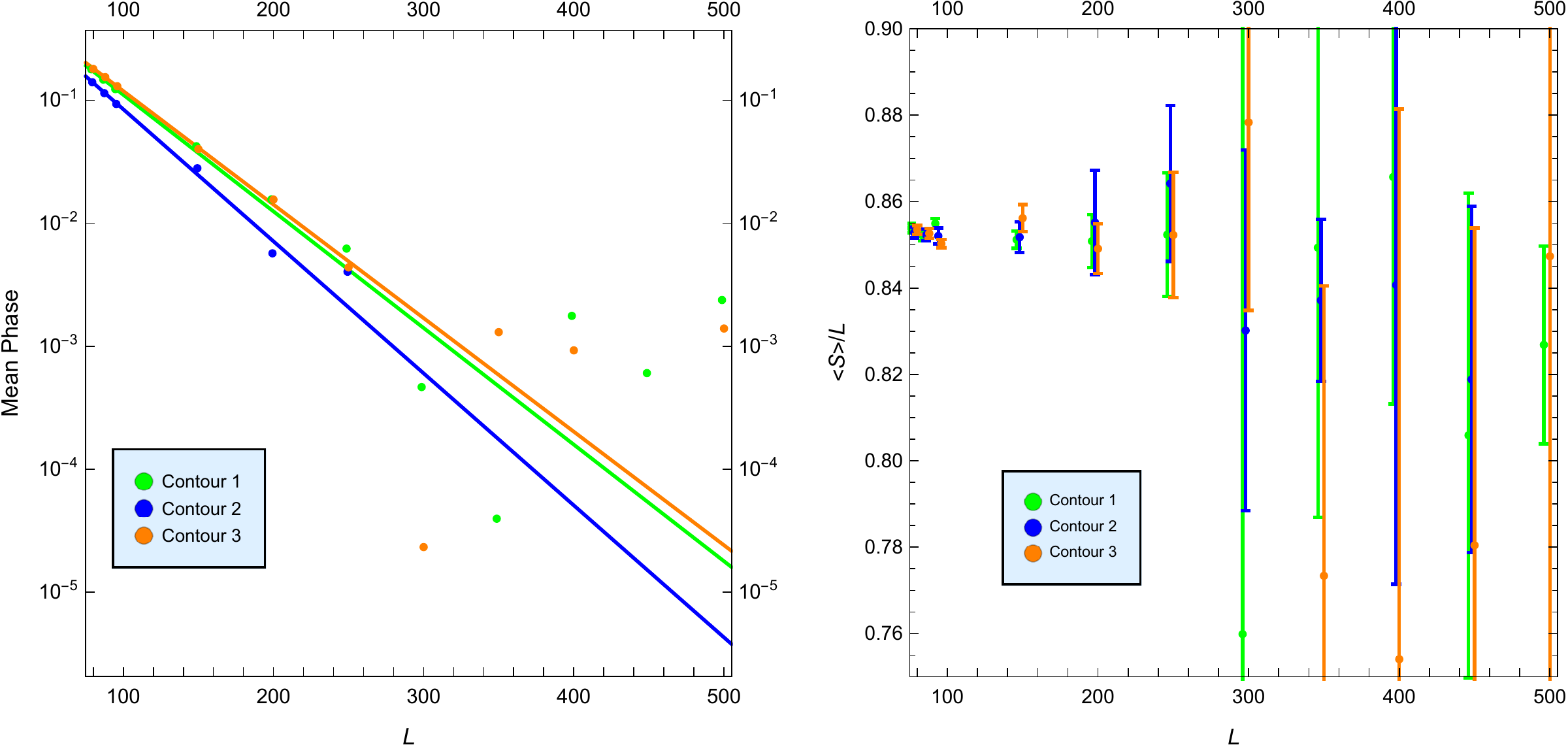}
\caption{The onset of the sign problem for contours 1,2,3. Left: The mean phase as a function of lattice size on a logarithmic scale
together with the previously obtained linear fits. Right: The action density as a function of lattice size.
The various contours are colour marked. Different contours are separated horizontally for clarity.}
\label{fig:phaseActionLarge}
\end{center}
\end{figure}

\begin{figure}[ht]
\begin{center}
  \includegraphics[width=5.95in]{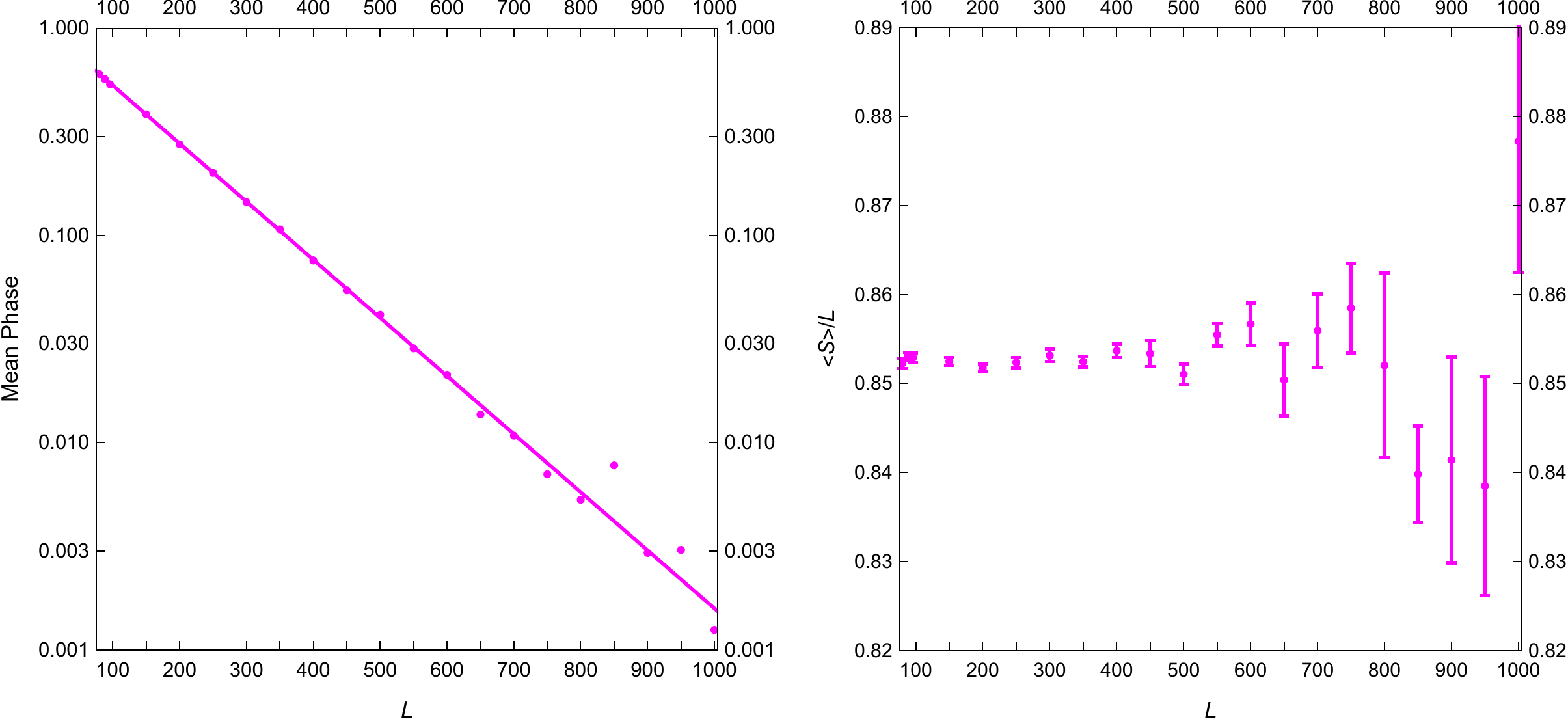}
\caption{The onset of the sign problem for contour 4. Left: The mean phase as a function of lattice size on a logarithmic scale
together with the previously obtained linear fit. Right: The action density as a function of lattice size.}
\label{fig:phaseActionLarge5}
\end{center}
\vspace{-6mm}
\end{figure}

In order to examine the suggested rule of thumb, we simulated contour 1,2,3
for $L=150,200,...,500$. The results are shown in fig.~\ref{fig:phaseActionLarge} together with the previously found fit.
The exponential decrease of the mean phase factor continues until about $L=200$,
where it begins to fluctuate. In some cases the mean phase becomes negative and so it cannot be presented on the logarithmic scale used.
Around the same value of $L$ the action density begins to fluctuate and to develop large standard errors. Yet another (unplotted) consequence
of the emergence of the sign problem is the appearance of large values of the imaginary part of the observables.
All that illustrates our prediction for this case.
Next, we examine the same predictions for contour 4 (we do not examine contour 5, for which we do not have a fast algorithm).
The results are presented in fig.~\ref{fig:phaseActionLarge5}. The behaviour is essentially the same and we conclude that the 
rule of thumb suggested holds in this case as well.
Such estimates can be very useful, since they can enable one to understand which $(p,q)$ order to pick for obtaining reliable results
on a given large lattice, by simulating other contours on much smaller lattices.

\subsection{Resolving the numerical instability}
\label{sec:resultsInstability}

As mentioned above, contour 4 suffers from numerical instability when simulated using
the fast algorithm that relies on~(\ref{newM}).
Four possible resolutions to the stability problem where offered at the end of section~\ref{sec:Jacobian}.
To these options we add ``option 0'', of completely ignoring the problem. We rewrite our options below and refer again
to the end of section~\ref{sec:Jacobian} for details:
\begin{enumerate}
\setcounter{enumi}{-1}
\item Do nothing.
\item Use LU decomposition; this is exact and numerically robust but slow.
\item Change the parameters to avoid the problem.
\item Leave the contour of one specific point undeformed.
\item Use the approximate form for the Jacobian~(\ref{approxM}).
\end{enumerate}
We examined the same observables as above for all these options. For option 2, we looked for other local maxima
(in the space of parameters) of the mean phase factor on a small lattice. In light of the form obtained for the parameters
defining the original contour 4 we considered only the 4-parameter
family of $U(1)$-invariant forward nearest neighbour
contours.
One of the local maxima we found did not suffer from numerical instability. The parameters of this contour are
\begin{equation}
\begin{aligned}
& a_{000001}=0.82\,,\qquad\qquad\qquad a_{000100}=1.27\,,\\
& b_{000002}=b_{002000}=0.10\,,\qquad b_{000200}=b_{200000}=0.40\,.
\end{aligned}
\end{equation}
We use these parameters for defining option 2 in what follows.
For option 3, we verified that no anomalous behaviour is obtained at the special point.

\begin{figure}[ht]
\begin{center}
  \includegraphics[width=5.9in]{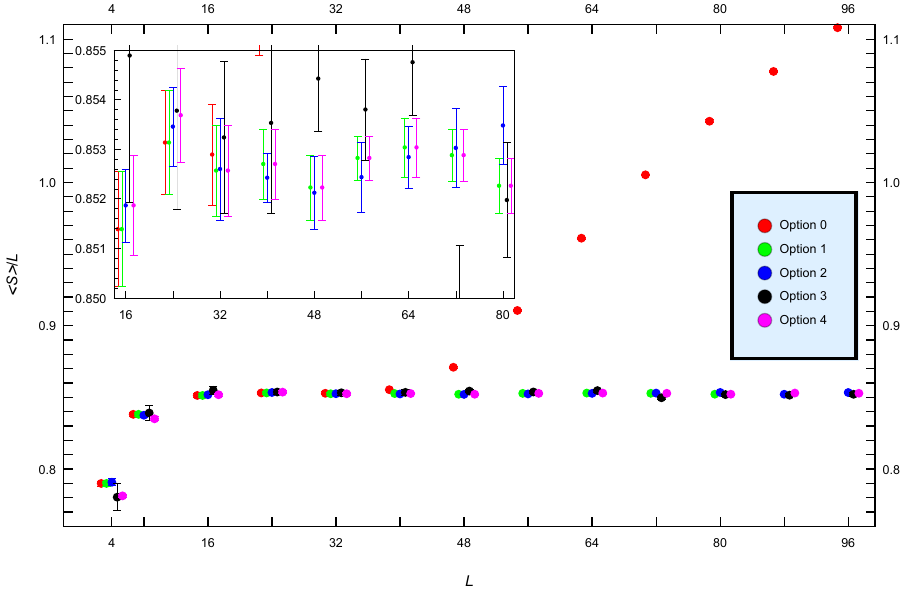}
\caption{The action density for the different options that can be used for resolving the numerical stability of contour 4.
The various options are colour marked.
Ignoring the problem (red) leads to an erroneous increase of the action density beyond $L\simeq 40$.
All other options lead to similar and consistent results.
As can be seen in the inset, in which the $y$ coordinate is magnified, for $L>L_c\equiv 24$ option 4 (using the approximate
expression for the Jacobian) produces similar results to the ones obtained by option 1 (using the exact algorithm).
In fact, the two options produce in this range identical results, within our working precision.}
\label{fig:inset}
\end{center}
\vspace{-6mm}
\end{figure}

In fig.~\ref{fig:inset} the results for the action obtained by the different options are compared.
We see that the ``approximate'' option 4, based on~(\ref{approxM}), is in fact exact beyond a given small critical value $L_c$.
Hence, in the previous subsection, we used the exact slow algorithm up to $L_c$ and
the ``approximate'' one from this value on.
One can still wonder why is the approximation so good and whether this would necessarily remain the case for all values of $L>L_c$.
It turns out that, within our working precision, this approximation is exact throughout this range.
The reason for that can be traced to the values that can be obtained
by the matrices $S_k$~(\ref{Sk}), from which the matrix $M$ is constructed. It is hard to evaluate these values analytically,
but we found out that throughout the run, the norm of each row or column vector, within the $S_k$,
never gets values above 0.631.
Then, using Cauchy-Schwarz inequality, the absolute value of all entries of $S_k S_{k+1}$ cannot exceed $0.631^2 = 0.398$.
Then, the entries of $S_k S_{k+1} S_{k+2} S_{k+3}$ cannot exceed $2\times 0.398^2=0.317$.
Continuing this way, one can deduce that as $L$ grows the possible maximal absolute value of the entries of $M$ becomes smaller and smaller,
until for some critical value $L_c$ it gets below our precision threshold.
Further increasing $L$ can only improve the approximation.
Note that the numbers written above can only be given for fixing
an upper bound on $L_c$. From the simulations we actually find that it is lower than this upper bound.

\begin{figure}[ht]
\begin{center}
  \includegraphics[width=5.6in]{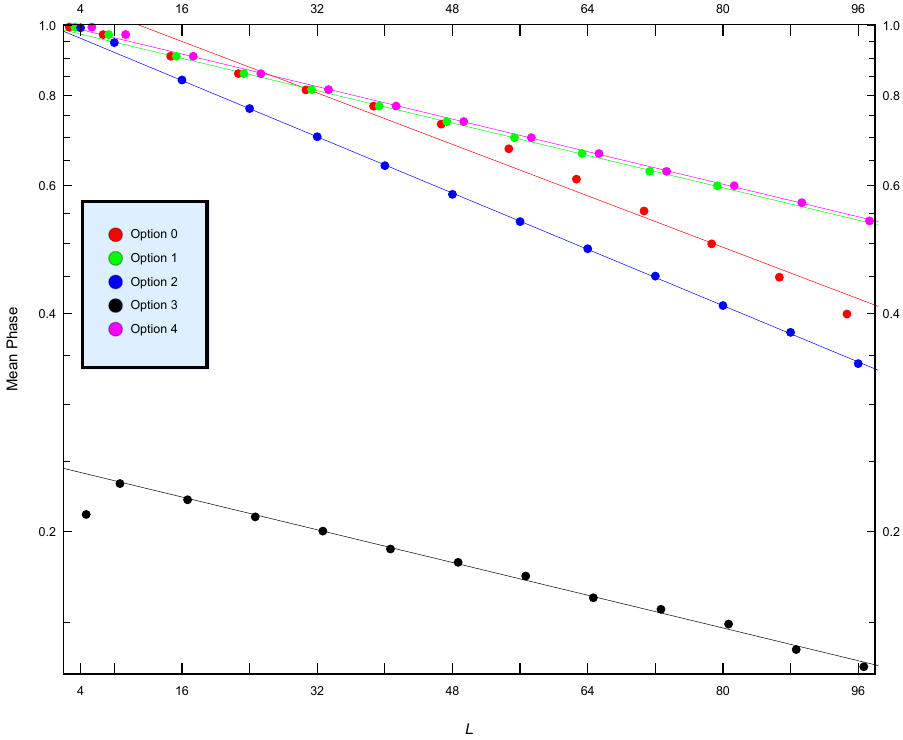}
\caption{The mean phase as a function of lattice size on a logarithmic scale for the different
options for resolving the numerical stability of contour 4 together with their linear least square fits.
The various options are colour marked and are separated horizontally for clarity.}
\label{fig:phases5f}
\end{center}
\vspace{-6mm}
\end{figure}

While using option 4 resolves the problem in the case at hand, it is still worth comparing the mean phase in all the cases,
since in the general case the actual numbers might be different and it is important to know how the other options could be of help.
In fig.~\ref{fig:phases5f} the mean phase as a function of lattice size is presented for the various options.
We see that the phase of option 0 does not decay exponentially in the range in which this is expected to happen,
i.e. the fit in this case is not particularly convincing.
This is another indication that something is wrong with this approach.
We use again the rule of thumb for predicting when the sign problem would kick in for the various other cases.
Extrapolating the numerical fits drawn in fig.~\ref{fig:phases5f} suggest that while options 1 and 4 (the ``contour 4''
of the previous subsection) are expected to give reliable results up to about $L=700$, option 2 should be reliable
up to about $L=400$, while option 3, which appears to be much lower than all the other ones in the plot, would
give consistent results up to about $L=500$. The reason for this relative success of this approach is that,
as expected, the slope in this case is almost identical to the slope of the usual contour 4 ($-0.00645$ for contour 4
as compared to $-0.00650$ for option 3 on the logarithmic scale). The difference between the two cases is related to
the large constant phase obtained from the one point whose contour was not deformed.
We deduce that in cases in which one cannot claim, as we do here, that the approximation~(\ref{approxM}) is exact,
treating one point differently from the others would probably be the best approach.
Moreover, one can easily increase the initial low phase in this case by allowing a deformation of the contour in the special
point that would be strictly local (that is, without a dependence on any nearest neighbour).

\section{Discussion}
\label{sec:conc}

In this work we illustrated, using a toy model, a viable method for addressing the sign problem.
The method relies on several simple principles, sketched in the introduction (section~\ref{sec:intro}).
We demonstrated that with a relatively low number of configurations (300,000) and with a small
number of free parameters, one can reliably simulate lattices of not too small size. We also
suggested how to generalise our construction in a way that should give reliable lattice simulations
also on quite large lattices. While the generalisation to other physical systems would require
repeating the steps used in the paper in order to construct an adequate ansatz, this construction
is relatively simple in principle and presumably can be performed for various systems.
The parameters describing the suggested contours depend on the values of the lattice parameters,
but their small number together with the ability to find them on a small lattice, suggest
that they can be easily found at the beginning of a simulation.

Our method leads to a significant improvement in running times. While we estimated in the introduction
that at $L=96$ the running time for contour 0 that suffers from a sign problem would be at least a couple of months
on a standard laptop, our actual running time was much shorter and led to more accurate results:
Without imposing the fast evaluation of the Jacobian, described in section~\ref{sec:Jacobian}, the running
time for $L=96$ was about one week, while using these results it reduced to about two minutes!
Results at $L=1000$ were obtained within about 20 minutes. Longer running times on fast computers
would enable obtaining results for very large lattices.

While the introduction of a fast algorithm for simulating forward nearest neighbours certainly helps, one could still worry
about the fact that only one nearest neighbour could be included. This is not quite so. One could extend
our expressions and allow the $y_k$ and $\zeta_k$ to depend on the fields at $k,k+1,k+2,...$, etc.
This would lead to almost upper-triangular matrices. The evaluation of the determinant could then be
performed either by setting some of the contours to the trivial ones, or by generalising the ideas
presented in section~\ref{sec:Jacobian}. In this way the approach can be extended to work also for
terms induced by the secondary interaction of next to nearest neighbours, etc.

Another possible reservation would concern the generalisation to higher dimensions:
It seems that when lattice points are coupled to neighbours from different space-time directions
one would not be able to obtain the almost upper-block-triangular matrix needed for the fast evaluation
of the Jacobian. Again, there are several ways out:
\begin{enumerate}
\item At any number of dimensions the 0 direction is special, since this is the direction related to
the chemical potential. Including only nearest neighbour terms in this direction would already take care
of the most significant contribution to the sign problem.
\item One could generalise the prescription of option 3 of subsection~\ref{sec:resultsInstability}, such that there would be no
deformations of the contour for lattice points that lie along specific co-dimension 1 hyper-surfaces
(one could allow some restricted deformations of the contour at these points). This would give an upper-block-triangular matrix.
There is a drawback to this approach: Since the number of points in the co-dimension one surface goes like $L^{d-1}$,
neglecting to deform them could lead to a significant sign problem by itself. Thus, this is probably not the best
option for $d > 1$.
\item One could split the lattice points such that some points would be coupled to neighbours
in the 0 direction and some others to neighbours in other directions, in a way that would still
produce an almost upper-block-triangular matrix. The approximate continuity of the field on the lattice
(enforced by the kinetic term) would induce an effective coupling of all fields in all directions.
\item One could generalise the method introduced here such that some non-locality would be allowed,
but in a way that could still be evaluated using a fast algorithm.
\item One could rewrite the action before summation such that the points near the boundary would be
treated differently. Then, when we impose the requirement that the terms vanish even before summation,
different equations would be obtained for these points, in a way that is consistent with the forward nearest
neighbour prescription.
\item There is always the option of using a slow algorithm for evaluating the Jacobian. For some systems
there is a bottleneck of $O(L^4)$ even without a sign problem, coming, e.g. from a fermionic determinant.
For such systems there is no reason to look for fast algorithms for the evaluation of the Jacobian.
Moreover, if all else fails for a system without such a bottleneck, one could still use the slow algorithm,
since while being slow compared to the $O(L)$ algorithm, it is still polynomial with not a very high power.
\end{enumerate}
We currently examine variants of all these possibilities and hope to describe the results in a future work.

Another important observation regarding generalisations to higher dimensions is the following one:
Our approach is based on an expansion around $\alpha=0$. In light of the definition of $\alpha$~(\ref{alphaDef})
we interpreted this as an expansion around the infinite mass limit. However, one could also think of
the expansion as being around the $d\rightarrow\infty$ limit. Hence, it would be natural to expect that
for a fixed number of lattice points and fixed values of the parameters, our approach would work
better for large $d$ than for small $d$. In particular, we expect that for the current model
our approach would deal better with the sign problem for $d=2,3,4$ than for the current $d=1$ case.
Given the fact that we could get up to almost 1,000 lattice points before the sign problem kicks in
at $d=1$, we believe that we would be able to get to about $10^4=10,000$ lattice points
at $d=4$, as in~\cite{Aarts:2008wh}, where the complex Langevin method was employed.
Moreover, in the current paper we considered only the first order in the $\alpha$ expansion and generalisations
of its form. Going to the second order is straightforward. It could further improve the success of the approach.
We currently examine these issues as well.

As mentioned, generalisation of the approach to local theories with more distant neighbours is simple.
However, in the case of a non local theory, e.g. a bosonic theory obtained after the integration of fermions,
implementation of the approach would be less straightforward. Since the implementation of the approach
is anyway model dependent, we refrain here from offering particular directions for this case and leave
this, very important, question for future work.

\section*{Acknowledgements}

We would like to thank Kouji Kashiwa, Scott Lawrence, Yuto Mori, Akira Ohnishi, and Benjamin Svetitsky for discussions.
The research of M.~K. was supported by the Israel Science Foundation (ISF), grant No. 244/17
and by EU FP7 IRSES program STREVCOMS, grant no. PIRSES-2013-612669.


\bibliography{bib}

\providecommand{\href}[2]{#2}\begingroup\raggedright\begin{thebibliography}{10}

\bibitem{deForcrand:2010ys}
P.~de~Forcrand, {\it {Simulating QCD at finite density}},  {\em PoS} {\bf
  LAT2009} (2009) 010, [\href{http://xxx.lanl.gov/abs/1005.0539}{{\tt
  1005.0539}}].

\bibitem{Aarts:2015tyj}
G.~Aarts, {\it {Introductory lectures on lattice QCD at nonzero baryon
  number}},  {\em J. Phys. Conf. Ser.} {\bf 706} (2016), no.~2 022004,
  [\href{http://xxx.lanl.gov/abs/1512.05145}{{\tt 1512.05145}}].

\bibitem{Medina:2017xbn}
L.~Medina and M.~C. Ogilvie, {\it Simulation of scalar field theories with
  complex actions},  \href{http://xxx.lanl.gov/abs/1712.02842}{{\tt
  1712.02842}}.

\bibitem{Bender:1998ke}
C.~M. Bender and S.~Boettcher, {\it Real spectra in {non-Hermitian Hamiltonians
  having PT} symmetry},  {\em Phys.Rev.Lett.} {\bf 80} (1998) 5243--5246,
  [\href{http://xxx.lanl.gov/abs/physics/9712001}{{\tt physics/9712001}}].

\bibitem{Bender:2007nj}
C.~M. Bender, {\it {Making sense of non-Hermitian Hamiltonians}},  {\em Rept.
  Prog. Phys.} {\bf 70} (2007) 947,
  [\href{http://xxx.lanl.gov/abs/hep-th/0703096}{{\tt hep-th/0703096}}].

\bibitem{Bursa:2014oza}
F.~Bursa and M.~Kroyter, {\it Lattice string field theory: The linear dilaton
  in one dimension},  {\em JHEP} {\bf 10} (2014) 74,
  [\href{http://xxx.lanl.gov/abs/1405.5089}{{\tt 1405.5089}}].

\bibitem{Witten:1986cc}
E.~Witten, {\it Noncommutative geometry and string field theory},  {\em Nucl.
  Phys.} {\bf B268} (1986) 253.

\bibitem{Metropolis:1953am}
N.~Metropolis, A.~W. Rosenbluth, M.~N. Rosenbluth, A.~H. Teller, and E.~Teller,
  {\it {Equation of state calculations by fast computing machines}},  {\em J.
  Chem. Phys.} {\bf 21} (1953) 1087--1092.

\bibitem{Gavai:2003mf}
R.~V. Gavai and S.~Gupta, {\it {Pressure and nonlinear susceptibilities in QCD
  at finite chemical potentials}},  {\em Phys. Rev.} {\bf D68} (2003) 034506,
  [\href{http://xxx.lanl.gov/abs/hep-lat/0303013}{{\tt hep-lat/0303013}}].

\bibitem{deForcrand:2002hgr}
P.~de~Forcrand and O.~Philipsen, {\it {The QCD phase diagram for small
  densities from imaginary chemical potential}},  {\em Nucl. Phys.} {\bf B642}
  (2002) 290--306, [\href{http://xxx.lanl.gov/abs/hep-lat/0205016}{{\tt
  hep-lat/0205016}}].

\bibitem{DElia:2002tig}
M.~D'Elia and M.-P. Lombardo, {\it {Finite density QCD via imaginary chemical
  potential}},  {\em Phys. Rev.} {\bf D67} (2003) 014505,
  [\href{http://xxx.lanl.gov/abs/hep-lat/0209146}{{\tt hep-lat/0209146}}].

\bibitem{Aarts:2013lcm}
G.~Aarts, {\it {Complex Langevin dynamics and other approaches at finite
  chemical potential}},  {\em PoS} {\bf LATTICE2012} (2012) 017,
  [\href{http://xxx.lanl.gov/abs/1302.3028}{{\tt 1302.3028}}].

\bibitem{Aarts:2009uq}
G.~Aarts, E.~Seiler, and I.-O. Stamatescu, {\it {The Complex Langevin method:
  When can it be trusted?}},  {\em Phys. Rev.} {\bf D81} (2010) 054508,
  [\href{http://xxx.lanl.gov/abs/0912.3360}{{\tt 0912.3360}}].

\bibitem{Aarts:2011ax}
G.~Aarts, F.~A. James, E.~Seiler, and I.-O. Stamatescu, {\it {Complex
  Langevin}: Etiology and diagnostics of its main problem},  {\em Eur. Phys.
  J.} {\bf C71} (2011) 1756, [\href{http://xxx.lanl.gov/abs/1101.3270}{{\tt
  1101.3270}}].

\bibitem{Nagata:2016vkn}
K.~Nagata, J.~Nishimura, and S.~Shimasaki, {\it {Argument for justification of
  the complex Langevin method and the condition for correct convergence}},
  {\em Phys. Rev.} {\bf D94} (2016), no.~11 114515,
  [\href{http://xxx.lanl.gov/abs/1606.07627}{{\tt 1606.07627}}].

\bibitem{Guralnik:2007rx}
G.~Guralnik and Z.~Guralnik, {\it Complexified path integrals and the phases of
  quantum field theory},  {\em Annals Phys.} {\bf 325} (2010) 2486--2498,
  [\href{http://xxx.lanl.gov/abs/0710.1256}{{\tt 0710.1256}}].

\bibitem{Witten:2010cx}
E.~Witten, {\it Analytic continuation of {Chern-Simons} theory},  {\em AMS/IP
  Stud. Adv. Math.} {\bf 50} (2011) 347--446,
  [\href{http://xxx.lanl.gov/abs/1001.2933}{{\tt 1001.2933}}].

\bibitem{Cristoforetti:2012su}
M.~Cristoforetti, F.~Di~Renzo, and L.~Scorzato, {\it New approach to the sign
  problem in quantum field theories: {High density QCD on a Lefschetz
  thimble}},  {\em Phys.Rev.} {\bf D86} (2012) 074506,
  [\href{http://xxx.lanl.gov/abs/1205.3996}{{\tt 1205.3996}}].

\bibitem{Alexandru:2016lsn}
A.~Alexandru, G.~Başar, P.~F. Bedaque, G.~W. Ridgway, and N.~C. Warrington,
  {\it {Fast estimator of Jacobians in the Monte Carlo integration on Lefschetz
  thimbles}},  {\em Phys. Rev.} {\bf D93} (2016), no.~9 094514,
  [\href{http://xxx.lanl.gov/abs/1604.00956}{{\tt 1604.00956}}].

\bibitem{Alexandru:2015xva}
A.~Alexandru, G.~Başar, and P.~Bedaque, {\it {Monte Carlo algorithm for
  simulating fermions on Lefschetz thimbles}},  {\em Phys. Rev.} {\bf D93}
  (2016), no.~1 014504, [\href{http://xxx.lanl.gov/abs/1510.03258}{{\tt
  1510.03258}}].

\bibitem{Alexandru:2016ejd}
A.~Alexandru, G.~Başar, P.~F. Bedaque, G.~W. Ridgway, and N.~C. Warrington,
  {\it {Monte Carlo calculations of the finite density Thirring model}},  {\em
  Phys. Rev.} {\bf D95} (2017), no.~1 014502,
  [\href{http://xxx.lanl.gov/abs/1609.01730}{{\tt 1609.01730}}].

\bibitem{Bedaque:2017epw}
P.~F. Bedaque, {\it A complex path around the sign problem},  {\em EPJ Web
  Conf.} {\bf 175} (2018) 01020,
  [\href{http://xxx.lanl.gov/abs/1711.05868}{{\tt 1711.05868}}].

\bibitem{Aarts:2008wh}
G.~Aarts, {\it Can stochastic quantization evade the sign problem? {The}
  relativistic {Bose} gas at finite chemical potential},  {\em Phys. Rev.
  Lett.} {\bf 102} (2009) 131601,
  [\href{http://xxx.lanl.gov/abs/0810.2089}{{\tt 0810.2089}}].

\bibitem{Mori:2017pne}
Y.~Mori, K.~Kashiwa, and A.~Ohnishi, {\it {Toward solving the sign problem with
  path optimization method}},  {\em Phys. Rev.} {\bf D96} (2017), no.~11
  111501, [\href{http://xxx.lanl.gov/abs/1705.05605}{{\tt 1705.05605}}].

\bibitem{Mori:2017nwj}
Y.~Mori, K.~Kashiwa, and A.~Ohnishi, {\it {Application of a neural network to
  the sign problem via the path optimization method}},  {\em PTEP} {\bf 2018}
  (2018), no.~2 023B04, [\href{http://xxx.lanl.gov/abs/1709.03208}{{\tt
  1709.03208}}].

\bibitem{Alexandru:2018fqp}
A.~Alexandru, P.~Bedaque, H.~Lamm, and S.~Lawrence, {\it Finite-density {Monte
  Carlo} calculations on sign-optimized manifolds},
  \href{http://xxx.lanl.gov/abs/1804.00697}{{\tt 1804.00697}}.

\bibitem{Alexandru:2018ddf}
A.~Alexandru, P.~F. Bedaque, H.~Lamm, S.~Lawrence, and N.~C. Warrington, {\it
  Fermions at finite density in (2+1)d with sign-optimized manifolds},
  \href{http://xxx.lanl.gov/abs/1808.09799}{{\tt 1808.09799}}.

\bibitem{Lawrence:2018mve}
S.~Lawrence, {\it Beyond thimbles: Sign-optimized manifolds for finite
  density},  in {\em {36th International Symposium on Lattice Field Theory
  (Lattice 2018) East Lansing, MI, United States, July 22-28, 2018}}, 2018.
\newblock \href{http://xxx.lanl.gov/abs/1810.06529}{{\tt 1810.06529}}.

\end{thebibliography}\endgroup

\vfill\eject

\end{document}